\documentclass{article}
\usepackage{ijcai21}

\usepackage{times}

\usepackage{soul}
\usepackage{url}
\usepackage[hidelinks]{hyperref}
\usepackage[utf8]{inputenc}
\usepackage[small]{caption}
\usepackage{graphicx}
\usepackage{amsmath}
\usepackage{booktabs}
\usepackage{subfigure}
\usepackage{balance} 
\usepackage{booktabs}       
\usepackage{amsfonts}       
\usepackage{nicefrac}       
\usepackage{microtype}      
\usepackage{amssymb}
\usepackage{appendix}
\usepackage{enumerate}
\usepackage{amsthm}
\usepackage{mathtools}
\usepackage{longtable}
\usepackage[ruled,linesnumbered,vlined]{algorithm2e}
\usepackage{makecell}
\usepackage{xcolor}
\usepackage{float}
\usepackage{wrapfig}
\usepackage{booktabs} 
\usepackage{thmtools,thm-restate}
\usepackage{enumitem}
\usepackage{cleveref} 

\theoremstyle{definition}

\allowdisplaybreaks[3]

\newcommand{\citet}[1]{\citeauthor{#1}~\shortcite{#1}}

\newcommand{\minisection}[1]{\vspace{5pt}\noindent{\bf #1.}}

\title{Deep Learning for Click-Through Rate Estimation}
\author{
    Weinan Zhang$^{\dag}$,
    Jiarui Qin$^{\dag}$,
    Wei Guo$^{\ddag}$,
    Ruiming Tang$^{\ddag}$,
    Xiuqiang He$^{\ddag}$\\
    \affiliations
    $^{\dag}$Shanghai Jiao Tong University ~~~~~~$^{\ddag}$Huawei Noah's Ark Lab\\
    \emails
    \{wnzhang, qinjr\}@apex.sjtu.edu.cn, 
    \{guowei67, tangruiming, hexiuqiang1\}@huawei.com
}

\begin{document}
\maketitle

\begin{abstract}
Click-through rate (CTR) estimation plays as a core function module in various personalized online services, including online advertising, recommender systems, and web search etc. From 2015, the success of deep learning started to benefit CTR estimation performance and now deep CTR models have been widely applied in many industrial platforms. 
In this survey, we provide a comprehensive review of deep learning models for CTR estimation tasks. First, we take a review of the transfer from shallow to deep CTR models and explain why going deep is a necessary trend of development. Second, we concentrate on explicit feature interaction learning modules of deep CTR models. Then, as an important perspective on large platforms with abundant user histories, deep behavior models are discussed. Moreover, the recently emerged automated methods for deep CTR architecture design are presented. Finally, we summarize the survey and discuss the future prospects of this field.
\end{abstract}

\section{Background}
Personalized services have become very important in various online information systems \cite{fan2006personalization}, such as item recommendation in e-commerce, auction in online advertising, page ranking in web search, to name a few.
Regarding the click as the representative behavior of user preference, click-through rate (CTR) estimation based on learning over the logged behavior data plays as a core function module in these personalization services \cite{agarwal2014laser}.

\begin{figure}[h]
    \centering
    \includegraphics[width=1\columnwidth]{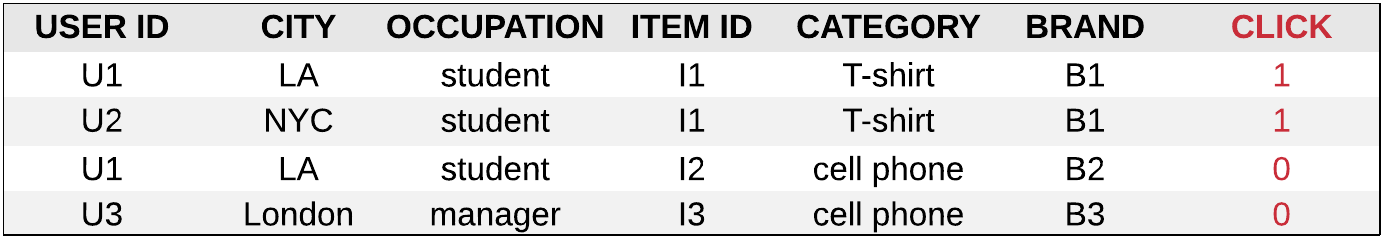}
    \caption{Multi-field categorical data for CTR estimation.}
    \label{fig:tabular-data}
\end{figure}


For CTR estimation tasks, the dataset is commonly represented as a table.
In the training pipeline of the CTR estimator, the historical instance set with a pre-defined size will be fetched, e.g., the previous 7-day logs. Then the feature engineering module will be executed, including feature normalization, discretization of continuous feature value, feature hashing and indexing etc. \cite{juan2016field}. 
Finally, each instance is in a multi-field categorical format while the corresponding label is binary (1 for click and 0 for non-click), 
with an example shown in Fig.~\ref{fig:tabular-data}.\footnote{Note that the scope of data format in this paper is focused on multi-field categorical data as shown in Fig.~\ref{fig:tabular-data}. Other content data such as text and image is not considered.}

As for the model training, the CTR estimation task is formulated as a binary classification problem, with the cross entropy loss function as
\begin{equation}\label{eq:ce-loss}
l(x,y,\Theta) = -y \log \sigma(f_\Theta(x)) - (1-y) \log (1 - \sigma(f_\Theta(x)))~,
\end{equation}
where $\sigma(z)=1/(1+\exp(-z))$ is the sigmoid function, $f_\Theta$ denotes the logit value function, parameterized by $\Theta$, $(x,y)$ denotes the instance with feature vector $x$ and label $y$. 

\begin{figure}[h]
    \centering
    \includegraphics[width=0.8\columnwidth]{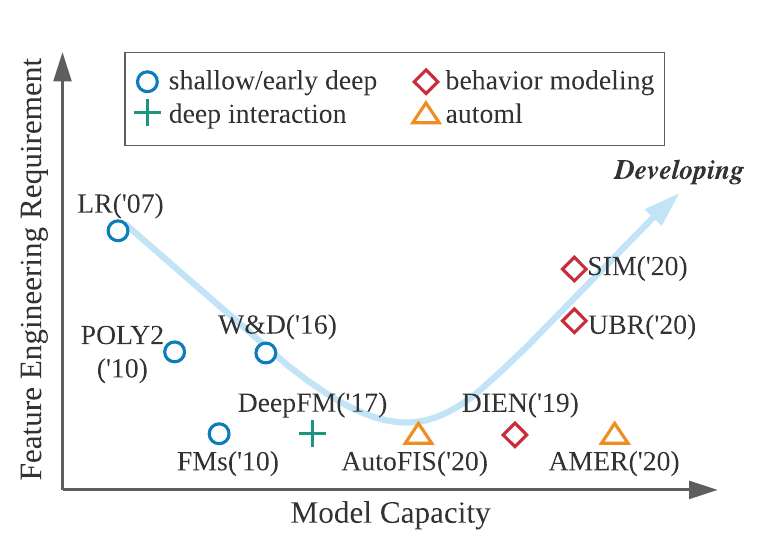}
    \vspace{-5pt}
    \caption{Development trend of CTR estimation models.}
    \label{fig:trend}
\end{figure}

We pick some representative models to demonstrate the development trend of CTR estimation models as illustrated in Fig.~\ref{fig:trend}.
The development of the models could be summarized into two aspects which are feature engineering complexity and model capacity.
For early CTR models, constrained by computing power, the major efforts have been made on designing better features by human with adopting simple models.
Later, more complicated models (with deep architectures and better modeling capacity) are introduced to liberate the complexity of human efforts for feature engineering. 
A more recent trend focuses again on feature engineering using some learnable methods, because it has come to a bottleneck of performance by solely designing more complicated deep models. 
Combining both complex models and learnable feature engineering is the new development direction.

\section{From Shallow to Deep CTR Models}\label{shallow_to_deep}

Dealing with such a binary classification task as in Eq.~(\ref{eq:ce-loss}), logistic regression (LR) \cite{richardson2007predicting} is the most basic model
with the advantage of high efficiency and ease for fast deployment. 
The logit value of LR is calculated as
\begin{equation}
    f_\Theta^{\text{LR}}(x) = \theta_0 + \sum_{i=1}^m x_i \theta_i~,
\end{equation}
where $\Theta=\{\theta_i\}_{i=0}^m$ and $m$ denotes the feature size.
However, many discriminative patterns for click prediction are combining features (or called cross features), e.g., user is a student (\textsc{Occupation:Student}) and location is LA (\textsc{City:LA}) will uplift the predicted CTR for a Disneyland ad (\textsc{Ad:Disneyland}), which corresponds to a third-order combining feature. 
Therefore, a straightforward way is to manually select and design many useful combining features, which requires enormous human efforts, as presented in Fig.~\ref{fig:trend}.

POLY2 \cite{chang2010training} assigns a weight to every second-order combining feature, 
which requires $O(m^2)$ parameter space. 
Unfortunately, the performance of POLY2 might be poor when the data is sparse. The key problem is that the parameters related to feature interactions cannot be correctly estimated in the sparse data where many features have never or seldom occurred together.

Another way to improve LR is to utilize the capacity of feature selection and combination of Gradient Boosting Decision Tree (GBDT) for learning feature interactions automatically \cite{GBFM}.
However, GBDT is hard to be trained in parallel and can only exploit a small fraction of possible feature interactions, which limit its performance and application in large-scale scenarios.

Factorization machine (FM) \cite{rendle2010factorization} assigns a $k$-dimensional learnable embedding vector $v_i$ to each feature $i$, which enables the model to explore the effectiveness of combining features in a flexible manner as
\begin{equation}
    f_\Theta^{\text{FM}}(x) = \theta_0 + \sum_{i=1}^m x_i \theta_i + \sum_{i=1}^m\sum_{j=i+1}^m x_i x_j v_i^\top v_j ~,
\end{equation}
where $\Theta=(\theta, v)$.
It is intuitive to see that if the combining feature $(i,j)$ is positively (negatively) correlated with $P(y=1)$, then the embedding inner product $v_i^\top v_j$ will be learned as positive (negative) automatically.
Further extensions of FM have been proposed.
The most notable extension is Field-aware FM (FFM) \cite{juan2016field}, which assigns multiple independent embeddings for a feature to explicitly model feature interactions with different fields. Despite the significant improvement over FM, FFM suffers from heavy parameter settings.
Besides, Gradient Boosting FM (GBFM) \cite{GBFM}, Higher-Order FM (HOFM) \cite{HOFM}, and Field-weighted FM (FwFM) \cite{pan2018field} are also proposed to improve FM.
By directly enumerating all the possible feature interactions, FM based models reduce the human involvement in feature engineering.

From 2015, the success of deep learning started to benefit CTR estimation performance via transferring the classic architectures or developing new ones.
The universal approximation property of neural networks \cite{cybenko1989approximation} and deep learning programming libraries with GPU computing stack make it possible to train deep neural network models to capture high-order feature interaction patterns and achieve better performance in CTR estimation.
With the vector representation of each sparse (or categorical) feature, the instance dense vector can be built by concatenating these vectors, which can be simply fed into a multi-layer perceptron (MLP) with a sigmoid output.
In industry, such an architecture is called DNN or sparse neural network (SNN). \citet{zhang2016deep} studied the parameter initialization of DNN for CTR estimation and found that the embedding vectors initialized by pre-training via a factorization machine (FNN) or a stacked auto-encoder can improve the prediction performance of DNN.
Wide \& Deep network \cite{cheng2016wide} is one of the earliest published deep models on CTR estimation, which adds the logit value of LR (wide part) and DNN (deep part) before feeding into the final sigmoid function as 
\begin{equation}
    f_\Theta^{\text{W\&D}}(x) = \theta_0 + \sum_{i=1}^m x_i \theta_i + \text{MLP}_\phi([v_1, v_2, \ldots, v_m]) ~,
\end{equation}
where $\Theta=(\theta, v, \phi)$.
Notice that manually designed cross features can also be included in the feature vector.
Thus the DNN part can be regarded as learning a residual signal of LR to further approach the label.
Moreover, DeepCross \cite{shan2016deep} extends residual network \cite{RESNET} for performing automatic feature interactions learning in an implicit way. 
Such early deep CTR models alleviate human efforts in feature engineering by incorporating simple MLPs. 

Although these early MLP-based deep models do achieve performance improvement, it may take a considerable effort to train a good model.
\citet{qu2018product} pointed out the insensitive gradient issue of DNN to capture discrete feature interactions and empirically showed that DNN with various sizes cannot well fit a POLY2 function.
\citet{rendle2020neural} found that it is more difficult for an MLP
to effectively learn the high-order combining feature patterns in CTR estimation tasks than a dot product in FM. 

Above findings suggest the explicit design of feature interaction learning in deep architectures as will be discussed in Section~\ref{sec:feat-interaction}. 
Later we will introduce the deep user behavior models in Section~\ref{sec:behavior} and the automated architecture search methods in Section~\ref{sec:automl}. 
One can see that (i) feature interaction learning mainly focuses on efficient pattern mining within an instance, (ii) user behavior modeling explores the dependency between multiple instances of a user, and (iii) automated architecture search methods aim to design above two kinds of deep models in a hands-free manner, which form the structure of the remaining part of this paper.



\section{Feature Interaction}\label{sec:feat-interaction}

\begin{figure*}[h]
    \centering
    \includegraphics[width=2.0\columnwidth]{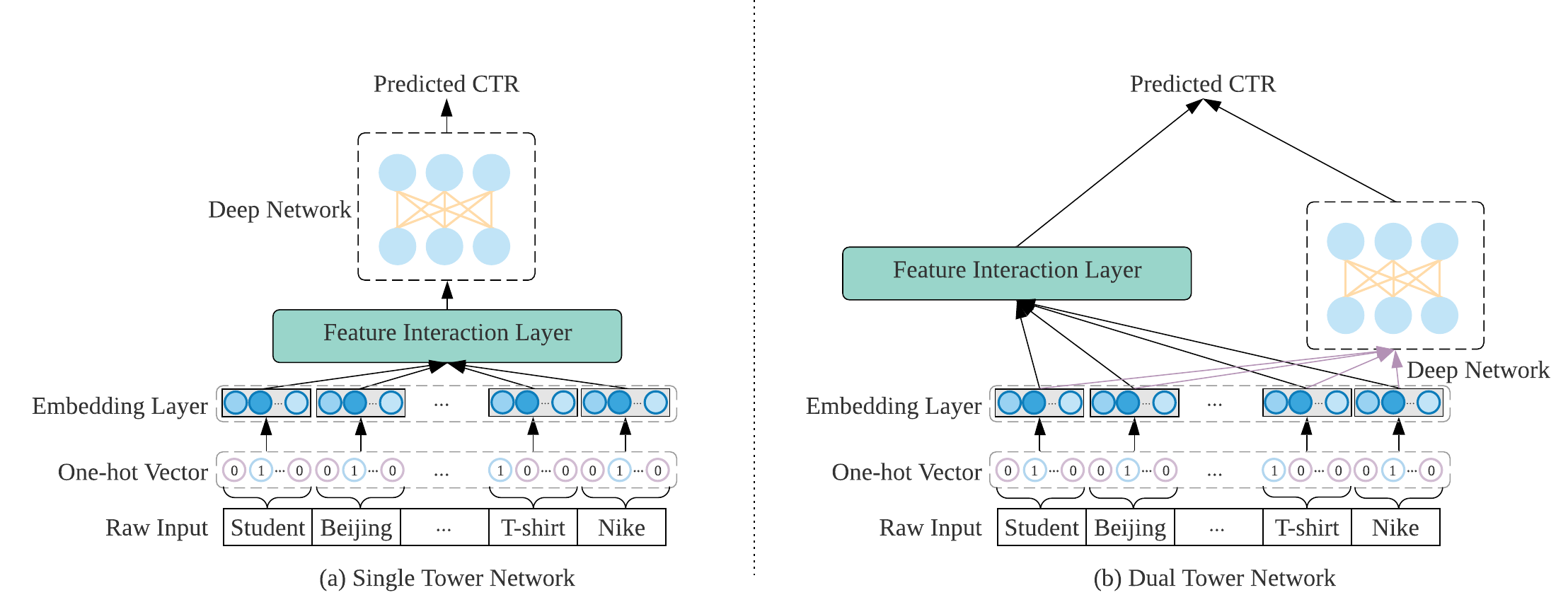}
    \caption{Illustration of single tower (left) and dual tower (right) deep CTR architectures.}
    \label{fig:single-dual}
   \vspace{-10pt}
\end{figure*}

As shown in Figure \ref{fig:single-dual},
the input multi-field categorical data is often represented as high-dimensional one-hot vectors.
An embedding layer is usually employed to compress these one-hot features into low-dimensional real-value vectors.
Feature interaction modeling, which indicates the combination relationships of multiple features, is the key to build a predictive model.
With the great power of feature representation learning, DNN is potential in automatic feature interactions modeling.
However, as pointed in Section \ref{shallow_to_deep}, it is difficult for a single DNN to learn high-order feature interactions effectively.
As a result, many works that incorporate explicitly feature interactions with DNN are proposed in recent years.
The feature interaction learning of existing models can be formulated as:
\begin{equation} \label{eq:fil}
    f_\Theta^{\text{FIL}}(x) = f_\psi([v_1, v_2, \ldots, v_{m}]) + \text{MLP}_\phi([v_1, \ldots, v_{m}]) ~,
\end{equation}
where $\Theta = (\phi,\psi)$ is the parameter set, $f_{\psi}$ is the explicitly feature interaction learning function and $\text{MLP}_\phi$ is an optional function that only dual tower models have.
We first give a detailed description of some existing operators for explicitly feature interaction learning, then we discuss the role of the DNN part in model architecture.  

\begin{figure*}[h]
    \centering
    \includegraphics[width=2.0\columnwidth]{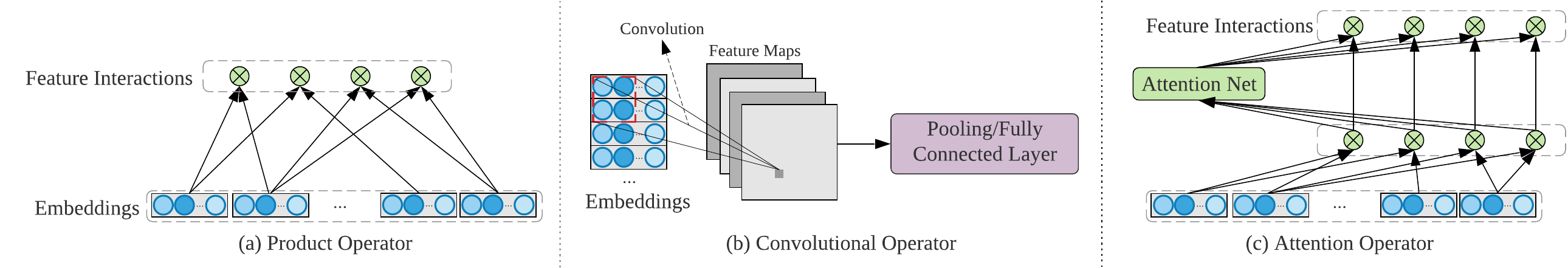}
    \caption{Illustration of three typical interaction operators.}
    \label{fig:interaction-funcs}
   \vspace{-10pt}
\end{figure*}

\subsection{Feature Interaction Operators}\label{sec:deep-fi}
There are multiple operators developed for explicitly feature interaction learning, which can be mainly classified into three categories, i.e., product operators, convolutional operators and attention operators.

\minisection{Product Operators}
To explicitly model the second-order feature interactions, Product-based Neural Network (PNN) \cite{IPNN} introduces a product layer between the embedding layer and deep network to model the feature interactions between different fields.
Specially, two variants are proposed, namely Inner Product-based Neural Network (IPNN) and Outer Product-based Neural Network (OPNN), which utilize an inner product layer or an outer product layer for feature interaction modeling, respectively.
Figure \ref{fig:interaction-funcs}(a) depicts the architecture of the inner product operation.
Experimental results show that adding product operations yields better convergence and improves the network's prediction ability. 
Neural Factorization machines (NFM) \cite{he2017neural2} finds that existing deep structures have difficulty in network training. Thus it proposes a bi-interaction layer between the embedding layer and deep network to model pairwise feature interactions.
By combining the linear feature interactions of the bi-interaction layer and the non-linear feature interactions of DNN,
NFM is much easier to train than existing deep models.
As the final function learned by DNN can be arbitrary, there is no theoretical guarantee on whether each order of feature interaction is modeled or not.
Cross Network \cite{wang2017deep} solves this problem using a cross network to apply feature crossing at each layer explicitly. 
Thus the order increases at each layer and is determined by layer depth.
Notice that Cross Network utilizes vector multiplication for feature crossing, which has a limited expressiveness, Cross Network V2 \cite{dcn-m} further replaces the cross vector in Cross Network into a cross matrix to make it more expressive. 
Inspired by Cross Network, Compressed Interaction Network (CIN) \cite{lian2018xdeepfm}, a more effective model, is proposed to capture the feature interactions of bounded orders.
To learn feature interactions better, Kernel Product Neural Network (KPNN) and Product-network In Network (PIN) utilize a kernel product and a micro-net architecture respectively to model feature interactions \cite{qu2018product}.

\minisection{Convolutional Operators}
Besides product operation for feature interaction modeling, Convolutional Neural Networks (CNN) and Graph Convolutional Networks (GCN) are also explored for feature interaction modeling.
Convolutional Click Prediction Model (CCPM) \cite{CCPM} performs convolution, pooling and non-linear activation repeatedly to generate arbitrary-order feature interactions, which is shown in Figure \ref{fig:interaction-funcs}(b).
However, CCPM can only learn part of feature interactions between adjacent features since it is sensitive to the field order.
Feature Generation Convolutional Neural Network (FGCNN) \cite{FGCNN} improves CCPM by introducing a recombination layer to model non-adjacent features.
It then combines the new features generated by CNN with raw features for final prediction.
FGCNN verifies that the generated features by CNN can augment the original feature space and reduce the optimization difficulties of existing deep structures.
Feature Interaction Graph Neural Networks (FiGNN) \cite{FIGNN} argues that existing deep models that use a simple unstructured combination of feature fields have a limited capacity to model sophisticated feature interactions. 
Inspired by the success of GCN, it treats the multi-field categorical data as a fully connected graph where different fields as graph nodes and interactions between different fields as graph edges, and then models feature interactions via graph propagation.

\minisection{Attention Operators}
Some works have tried to exploit the attention mechanism for feature interaction modeling in CTR estimation.
Attentional Factorization Machines (AFM) \cite{xiao2017attentional} improves FM by utilizing an additional attention network to enable feature interactions to contribute differently to the prediction.
The pairwise interacted vector of two features is fed into the attention network to calculate an attention score for this interacted vector at each step, as shown in Figure \ref{fig:interaction-funcs}(c).
Then a softmax function is applied to normalize these attention scores.
Finally, these attention scores are multiplied with the interacted vectors for final prediction.
Feature Importance and Bilinear feature Interaction NETwork (FiBiNET) \cite{FIBINET} extends the Squeeze-Excitation network (SENET) \cite{SENET} which has made great success in computer vision to learn the feature importance. It then uses a bilinear function to learn the feature interactions.
Inspired by the success of self-attention \cite{vaswani2017attention} in natural language processing, 
AutoInt \cite{AUTOINT} utilizes a multi-head self-attentive neural network with residual connections to explicitly model the feature interactions with different orders and can also provide explainable prediction via attention weights.
Interpretable CTR prediction model with Hierarchical Attention (InterHAt) \cite{li2020interpretable} combines a transformer network with multiple attentional aggregation layers for feature interactions learning,
which achieves high training efficiency with comparable performance and can explain the importance of different feature interactions.

\subsection{The Role of DNN in Deep CTR Models}
The feature interaction operators in Section~\ref{sec:deep-fi} can be utilized to construct single tower networks or dual tower networks according to the relative positions of DNN and feature interaction layer in the architecture.
Single tower models place feature interaction and deep network successively in the architecture, as illustrated in Fig.~\ref{fig:single-dual}(a). These models could effectively capture the high-order feature interactions but the signals of low-order feature interactions may vanish in the following DNN.
To better capture low-order feature interactions, dual tower networks are proposed. As shown in Fig.~\ref{fig:single-dual}(b), it places the feature interaction layer and DNN parallelly. The feature interaction layer is responsible for explicitly capturing low-order interactions while the high-order ones are captured implicitly by the DNN. Outputs of both modules are used to generate the final prediction. 
Single tower models like NFM \cite{he2017neural2} and PIN \cite{qu2018product} have a stronger modeling capacity with a more sophisticated network structure. But they usually suffer from the bad local minima and heavily rely on parameter initialization.
Wide \& Deep \cite{cheng2016wide}, DeepFM \cite{guo2017deepfm}, DCN \cite{wang2017deep}, DCN V2 \cite{dcn-m}, xDeepFM \cite{lian2018xdeepfm} and Autoint+ \cite{AUTOINT} are dual tower models. The DNN part can always be regarded as a supplementary to learn the residual signal of the feature interaction layer to approach the label, which yields stable training and the improved performance.

\section{User Behavior Modeling}\label{sec:behavior}
User behaviors contain crucial patterns of user interests.
Modeling user behaviors is becoming an essential topic of the CTR estimation task in recent years. 
In this section, we will give an elaborate review of the user behavior modeling literature. 

\begin{figure}[h]
    \centering
    \includegraphics[width=1\columnwidth]{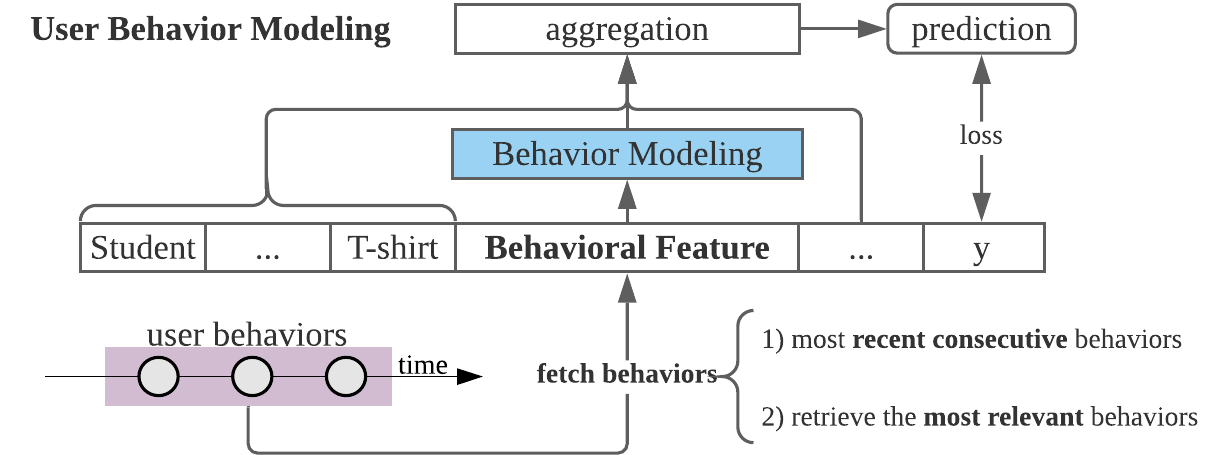}
    \vspace{-5pt}
    \caption{General framework of user behavior modeling.}
    \label{fig:ubm}
    \vspace{-5pt}
\end{figure}

The user behaviors are usually organized as a multi-value categorical feature, each value is a behavior identifier. 
The behavioral feature is usually sequential, sorted with the temporal order.
The item IDs and the corresponding item features (if any) form a behavior sequence.
The overall framework of user behavior modeling is illustrated in Fig.~\ref{fig:ubm}. In general, it could be summarized as Eq.~\eqref{eq:ubm}, where $x_m = \{b_1,b_2,\ldots,b_T\}$ is the multi-value behavioral feature, $T$ is the sequence length, $\Theta = (\phi,\psi)$ is the parameter set and $\mathcal{U}_{\psi}$ is the user behavior modeling function. Then the logit value is calculated as
\begin{equation} \label{eq:ubm}
    f_\Theta^{\text{UBM}}(x) =\text{MLP}_\phi([v_1, v_2, \ldots, v_{m-1}, \mathcal{U}_{\psi}(x_m)])~.
\end{equation}
The key point is to design an effective $\mathcal{U}_{\psi}$ function to learn a dense representation of the behavioral feature. After getting the behavioral representation, it is aggregated (concatenation) with other feature embeddings, and the aggregated features will be fed into an MLP to get the final prediction. 
The user behavior modeling algorithms could be categorized into three classes, i.e., attention based models, memory network based models and retrieval based models. There are also many user behavior modeling methods \cite{fang2020deep} in sequential recommendation task, but they are not included in this section because our paper concentrates on CTR estimation task.

\subsection{Attention based Models}
Deep Interest Network (DIN) \cite{zhou2018deep} is the first model that introduces attention mechanism in user behavior modeling of CTR estimation task. 
It utilizes an attention mechanism to attribute different historical behaviors with different weights according to the relatedness with the target item. 
As the user's latent interests evolve over time dynamically, it is vital to capture the temporal patterns inside the behavior sequence, which motivates the design of Deep Interest Evolution Network (DIEN) \cite{zhou2019deep}.
DIEN is a two-layer GRU (Gate Recurrent Unit) structure with the attention mechanism. The first GRU layer is responsible for interest extraction, and the second layer is used for modeling the interest evolving. AUGRU (GRU with attentional update gate) is proposed for the interest evolving layer, which uses the attention weights to update the states of GRU.
DIEN further incorporates an auxiliary loss that uses consecutive behavior to supervise the learning of hidden states at each step, making hidden states expressive enough to represent latent interest.
As self-attention \cite{vaswani2017attention} has shown promising performance gain in many applications, it is also used in user behavior modeling. Behavior Sequence Transformer (BST) \cite{chen2019behavior} directly incorporates multi-head attention layer as the sequential feature extractor to capture the dependency among behaviors. Deep Session Interest Network (DSIN) \cite{feng2019deep} splits user behaviors into multiple sessions. It uses self-attention with bias encoding to get accurate behavior representation inside each session and employs Bi-LSTM to capture the sequential relationship among historical sessions.
The above models have shown the efficacy of attention mechanism in user behavior modeling.

\subsection{Memory Network based Models}
As there is a massive amount of user behavior data accumulated on large e-commerce platforms \cite{ren2019lifelong}, it is of more importance to handle the very long behavior sequence and mine the patterns in more distant user history.
Models like DIN or DIEN is limited by time complexity, and they are not feasible in modeling very long sequence. Thus memory network based models are proposed with proper system designs.
\citet{ren2019lifelong} proposed Hierarchical Periodic Memory Network (HPMN) and the lifelong sequential modeling framework.
It uses a lifelong personalized memory to store the user interest representations, and HPMN is responsible for updating the memory states incrementally. HPMN is a multi-layer GRU architecture with different update frequencies at each layer. The upper layers update less frequently than the lower layers, thus HPMN could capture long-term yet multi-scale temporal patterns. 
With similar motivations, \citet{pi2019practice} designed a User Interest Center (UIC) module which decouples the time-consuming user behavior modeling procedure with the real-time prediction process. At the inference time, the user behavioral representation is directly obtained from UIC, which is calculated offline. UIC proposes a Multi-channel user Interest Memory Network (MIMN) to calculate the user behavioral representations as the user interests are usually multi-channeled.
Compared with HPMN, UIC provides a more systematic solution to the long sequence modeling from the industrial perspective, while HPMN is the first to incorporate the memory concept to handle long sequence modeling in the CTR estimation task.


\subsection{Retrieval based Models}
Despite the tremendous success of the current user behavior modeling solutions, there are still some drawbacks. All the above models use the most \textit{recent consecutive} behaviors and try to incorporate complex architectures to capture sequential patterns in longer sequences, which bring heavy burdens on the system overhead. 
User Behavior Retrieval (UBR4CTR) \cite{qin2020user} proposes a new framework that uses search engine techniques to retrieve the most relevant behaviors from the entire user behavior sequence. 
Instead of using a large number of consecutive behaviors, only a small number of the most relevant behaviors will be retrieved at each inference time. 
This retrieval approach not only solves the time complexity problem of handling very long sequences but also alleviates the noisy signals in long consecutive sequences because only the most relevant behaviors are input to the model.

UBR4CTR uses search engine techniques to retrieve the top-k relevant historical behaviors. 
The features of the inference target are selected to generate the query. The query generation procedure is parametric. Thus, it makes the retrieval process learnable, and it is optimized using REINFORCE algorithm.
Search-based Interest Model (SIM) \cite{qi2020search} is another retrieval based model which proposes hard search and soft search approaches. For the hard search, it uses predefined IDs such as user ID and category ID to build the index. 
As for the soft search, SIM retrieves relevant behaviors using their embeddings by local sensitive hashing (LSH).

Most of the user behavior models mentioned in this section are deployed in real-world online systems.
These user behavior modeling algorithms and systems contribute significantly to the development of CTR estimation techniques in terms of both economic returns and academic values.

\section{Automated Architecture Search}\label{sec:automl}

As some automatically designed networks achieve comparable performance to the human-developed models in computer vision, researchers of recommender systems also propose to devise automated methods for deep CTR architecture design. Depending on the different parts that such methods focus on, they can be categorized into three classes: (1) automatic design of flexible and adaptable embedding dimensions for individual features; (2) automatic selection or generation of effective feature interactions; (3) automatic design of network architectures.

\subsection{Embedding Dimension Search}

Embedding representation is a key factor to deep CTR models, as it is the major component of model parameters, which naturally has a high impact on the model performance. Therefore, some works optimize the embeddings by searching embedding dimensions for different features adaptively and automatically.

NIS~\cite{auto-embed-nis} and ESAPN~\cite{auto-embed-ESAPN} perform reinforcement learning (RL) to search for mixed feature embedding dimensions automatically. NIS~\cite{auto-embed-nis} first divides the universal dimension space into several blocks predefined by human experts and then applies RL to generate decision sequences on selecting such dimension blocks for different features. The reward function of the RL algorithm considers both the accuracy of the recommendation model and memory cost. ESAPN~\cite{auto-embed-ESAPN} also predefines a set of candidate embedding dimensions for different features. A policy network is utilized for each field to search the embedding sizes for different features dynamically. The policy network takes the frequency and current embedding size of a feature as the input, and outputs a decision of whether to enlarge the current dimension of this feature.

The above two methods select feature dimensions from a small discrete set of candidate dimensions, whereas DNIS~\cite{auto-embed-dnis} and PEP~\cite{auto-embed-pep} make the candidate dimensions continuous. More specifically, DNIS~\cite{auto-embed-dnis} introduces a binary indicator matrix for each feature block (features are grouped into feature blocks according to their frequency to reduce the search space) to indicate the existence of the corresponding dimension. A soft selection layer is then proposed to relax the search space of the binary indicator matrix to be continuous. 
Then a predefined threshold value is utilized to filter unimportant dimensions in the soft selection layer. However, this threshold value is hard to tune in practice and thus may lead to suboptimal model performance. Motivated by this observation, PEP~\cite{auto-embed-pep} prunes embedding parameters where the threshold values can be adaptively learned from data.

The works mentioned above perform a hard selection in the search space, i.e., only selecting one embedding dimension for each feature. On the contrary, soft selection strategy is proposed in AutoEmb~\cite{auto-embed-autoemb} and AutoDim~\cite{auto-embed-autodim}. The soft selection strategy sums over the embeddings of the candidate dimensions with learnable weights, where such weights are trained via differentiable search algorithms (e.g., DARTS). AutoDim allocates different embedding sizes for different feature fields, while AutoEmb searches different embedding sizes for individual features.

\subsection{Feature Interaction Search}

As stated earlier, feature interactions modeling is crucial in deep CTR models, thus automatically devising effective feature interactions is of high potential value.

AutoFIS~\cite{auto-fi-autofis} automatically identifies and then selects important feature interactions for factorization models. AutoFIS enumerates all the feature interactions and utilizes a set of architecture parameters to indicate the importance of individual feature interactions. These architecture parameters are optimized by gradient descent (inspired by DARTS~\cite{DARTS}) with GRDA optimizer~\cite{GRDA} to get a sparse solution, such that those unimportant feature interactions (i.e., with architecture parameters of zero values) will be filtered away automatically. However, AutoFIS restricts the interaction function to be the inner product.  

As stated in PIN~\cite{qu2018product}, a micro-network could be more effective than the inner product as the interaction function. Inspired by this observation, besides identifying important feature interactions, determining suitable interaction functions is also beneficial, which motivates SIF~\cite{auto-fi-sif} and AutoFeature~\cite{auto-fi-autofeature}.
SIF~\cite{auto-fi-sif} automatically devises suitable interaction functions for matrix factorization (considering only user and item identifier, without any other features) in different datasets. The search space of the interaction functions consists of micro space and macro space, where micro space refers to element-wise MLP, and macro space includes five linear algebra operations.
AutoFeature~\cite{auto-fi-autofeature} proposes to utilize a micro-network with different architectures to model feature interactions between each pair of fields. The architecture of each micro-network is searched from a search space with five pre-defined operations. The search process is implemented by an evolutionary algorithm with the Naive Bayes tree. 

However, neither AutoFIS nor AutoFeature can model high-order feature interactions, as they need to enumerate all the possible feature interactions beforehand. To avoid such inefficient enumeration, AutoGroup~\cite{liu2020autogroup} proposes to generate some groups of features, such that their interactions of a given order are effective. Each feature has an initial probability of 0.5 to be selected in any of such groups. Such probability values are parameterized and learned by Gumbel-Softmax trick~\cite{gumbel-softmax}, from the supervised signal of the CTR estimation task.

All the works mentioned above select or generate effective feature interactions for feature fields. 
BP-FIS~\cite{auto-fi-bpfis} is the first work to identify important feature interactions for different users by Bayesian variable selection, providing finer granularity of feature interaction selection than previous works. Specifically, a Bayesian generative model is built with a derived lower bound, which can be optimized by an efficient stochastic gradient variational Bayes method to learn the parameters.  

\subsection{Whole Architecture Search} 

The last category of works studies searching the whole architecture of deep CTR models.

AutoCTR~\cite{auto-nas-autoctr} designs a two-level hierarchical search space by abstracting representative structures in state-of-the-art CTR estimation architectures (namely, MLP, dot product and factorization machine) into virtual blocks, where such blocks are connected as a directed acyclic graph (DAG) in a similar way as in DARTS. The outer space consists of the connections among blocks, while the inner space is composed of the detailed hyperparameters in different blocks. The evolutionary algorithm is utilized as the search algorithm, with some optimizations on the efficiency.

AMER~\cite{auto-nas-amer} searches architectures to extract sequential representation from sequential features (i.e., user behaviors) and explores different feature interactions among non-sequential features automatically and simultaneously. On one hand, the search space of behavior modeling includes normalization, activation, and layer selection (e.g., convolution, recurrent, pooling, attention layers). Several architectures are randomly sampled for evaluation on the validation set, and only the top ones are selected for further evaluation. On the other hand, effective feature interactions are searched by progressively increasing the order of interactions. The candidate feature interactions are initialized by single non-sequential features and updated by interacting with all possible non-sequential features and keeping the best interactions with the highest validation performance.

\section{Summary and Future Prospects}
This paper provides a brief review of the development of deep learning models for CTR estimation tasks. With feature-interaction operators, deep models are more capable of capturing high-order combining feature patterns in the multi-field categorical data and yield better prediction performance. 
With attention mechanism, memory networks or retrieval-based approaches, the representation of a user behavior history can be effectively learned, which further improves the prediction performance.
As the deep architectures for CTR estimation can be various, some trials have been performed to automatically search the deep architectures to make the whole process hands-free. 
All the above progresses are achieved within the last five years.

Despite the fast development and great success of deep learning on CTR estimation,
we still note that there are some significant challenges in this field to be addressed.
\begin{itemize}[leftmargin=10pt]
    \item \textbf{Deep learning theory.} There are plenty of works on designing CTR models, but seldom works have focused on the deep learning theory of these models, including the sample complexity analysis, learning behaviors of feature interaction layers, gradient analysis etc.
    \item \textbf{Representation learning.} Just like other discrete data such as text and graph, it is reasonable to expect that the representation learning (or called pretraining) of the multi-field categorical data will largely improve the CTR prediction performance. However, to our knowledge, there has been rare existing work on this perspective so far.
    \item \textbf{Learning over multi-modal data.} In modern information systems, there are various multimedia elements of items or browsing environments. Thus it is of high potential to design CTR estimation models to deal with the feature interactions over multi-modal data. 
    \item \textbf{Strategical data processing.} The recent progress of strategies of user historical behavior fetching, e.g., user behavior retrieval and ranking, suggests the great potential of exploring the combining way of data processing with deep model design. Making such data processing learnable would be of high value for research. 

\end{itemize}

\section*{Acknowledgement}
Ruiming Tang is the corresponding author. Weinan Zhang is supported by ``New Generation of AI 2030'' Major Project (2018AAA0100900) and National Natural Science Foundation of China (61772333, 61632017). The work is also sponsored by Huawei Innovation Research Program.


\bibliographystyle{named}
\bibliography{deepctr}

\end{document}